\documentclass[journal]{IEEEtran}

\usepackage{ifpdf}

\usepackage{cite}

\ifCLASSINFOpdf
   \usepackage[pdftex]{graphicx}
\else
\fi
\usepackage[cmex10]{amsmath}
\usepackage{array}

\usepackage{mdwmath}
\usepackage{mdwtab}

\usepackage{eqparbox}

\usepackage[tight,footnotesize]{subfigure}

\usepackage{fixltx2e}
\usepackage{amsmath}
\usepackage{amssymb}
\usepackage{amsthm}
\begin{document}
%
\title{Development and Test of a $\mu$TPC Cluster Reconstruction for a Triple GEM Detector\\in Strong Magnetic Field}
%
%
%
\author{R.~Farinelli$^a$$^b$ *,M.~Alexeev$^c$, A.~Amoroso$^c$, F.~Bianchi$^c$, M.~Bertani$^d$, D.~Bettoni$^a$, N.~Canale$^a$$^b$, A.~Calcaterra$^d$, V.~Carassiti$^a$, S.~Cerioni$^d$, J.~Chai$^c$,  S.~Chiozzi$^a$, G.~Cibinetto$^a$, A.~Cotta Ramusino$^a$, F.~Cossio$^c$, F.~De~Mori$^c$, M.~Destefanis$^c$, T.~Edisher$^d$, F.~Evangelisti$^a$, L.~Fava$^c$, G.~Felici$^d$, E.~Fioravanti$^a$, I.~Garzia$^a$$^b$, M.~Gatta$^d$, M.~Greco$^c$, D.~Jing$^d$, L.~Lavezzi$^c$$^e$, C.~Leng$^c$, H.~Li$^c$, M.~Maggiora$^c$, R.~Malaguti$^a$, S.~Marcello$^c$, M.~Melchiorri$^a$, G.~Mezzadri$^a$$^b$, G.~Morello$^d$,S.~Pacetti$^f$, P.~Patteri$^d$, J.~Pellegrino$^c$, A.~Rivetti$^c$, M.~D.~Rolo$^c$, M.~Savrie'$^a$$^b$, M.~Scodeggio$^a$$^b$, E.~Soldani$^d$, S.~Sosio$^c$, S.~Spataro$^c$, L.~Yang$^c$.
\\ ~
\\ $^a$ INFN - Sezione di Ferrara, $^b$ University of Ferrara,$^c$ INFN - Sezione di Torino,\\ $^d$ INFN - Sezione di Frascati,  Physics dept., $^e$ IHEP, Beijing, $^f$ University of Perugia.
\\E-mail address: rfarinelli@fe.infn.it (R.~Farinelli), *Corresponding author.}%

\maketitle

\begin{abstract}
Performance of triple GEM prototypes has been evaluated by means of a muon beam at the H4 line of the SPS test area at CERN. The data from two planar prototypes have been reconstructed and analyzed offline with two clusterization methods: the center of gravity of the charge distribution and the micro~Time~Projection~Chamber ($\mu$TPC). 
GEM prototype performance evaluation, performed with the analysis of data from a TB, showed that two-dimensional cluster efficiency is above 95\% for a wide range of operational settings. Concerning the spatial resolution, the charge centroid cluster reconstruction performs extremely well with no magnetic field: the resolution is well below 100 $\mu m$ . Increasing the magnetic field intensity, the resolution degrades almost linearly as effect of the Lorentz force that displaces, broadens and asymmetrizes the electron avalanche. Tuning the electric fields of the GEM prototype we could achieve the unprecedented spatial resolution of 190 $\mu m$ at 1 Tesla. 
In order to boost the spatial resolution with strong magnetic field and inclined tracks a $\mu$TPC cluster reconstruction has been investigated. Such a readout mode exploits the good time resolution of the GEM detector and electronics to reconstruct the trajectory of the particle inside the conversion gap. Beside the improvement of the spatial resolution, information on the track angle can be also extracted. 
The new clustering algorithm has been tested with diagonal tracks with no magnetic field showing a resolution between 100 $\mu m$ and 150 $\mu m$ for the incident angle ranging from 10$^\circ$ to 45$^\circ$. Studies show similar performance with 1 Tesla magnetic field. This is the first use of a $\mu$TPC readout with a triple GEM detector in magnetic field. 
This study has shown that a combined readout is capable to guarantee stable performance over a broad spectrum of particle momenta and incident angles, up to a 1 Tesla magnetic field.
\end{abstract}

\begin{IEEEkeywords}
GEM, CGEM, $\mu$TPC, magnetic field, gas detector.
\end{IEEEkeywords}

%
\IEEEpeerreviewmaketitle

 \begin{figure}[!t]
 \centering
  \subfigure[]
    {    \includegraphics[width=1.1in]{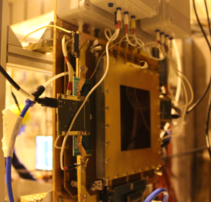}}
  \hspace{1.5mm}
  \subfigure[]
    {\includegraphics[width=2.1in]{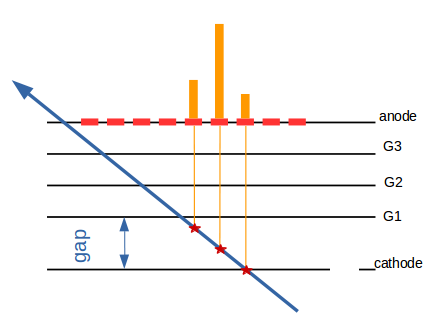}}
  \caption{a) Triple GEM detector instrumented during a test beam at CERN. b) Schematization of the triple GEM technology: the charge particle generate primary electron that drift to the anode. At each GEM the multiplication occurs.}
    \label{gemm}
  \end{figure}

\section{Introduction}
%
%
%
%
\IEEEPARstart{G}{as} detectors are instruments used to reveal the charged particles that pass throught them. The primary electron generated by the ionization of the gas are amplified and the electrical signal is collected to measure the spatial position of the particles. Compared to the first technology of gas detector, the Micro Pattern Gas Detector (MPGD) bypass the limits of the diffusion using a pitch/cell size of few hundred $\mu m$ that improves granularity and rate capability \cite{chinese}. Among the MPGD family, the Gas Electron Multiplier (GEM) technology invented by F.Sauli in 1997 \cite{sauli} exploits 50 $\mu m$ kapton foils covered on both sides by 5 mm of copper, and pierced with 50 $\mu m$ holes to amplify the electron signal by applying an electric field up to 10$^5$ kV/cm between the two faces. The signal is then readout by means of external strips of pads. A triple GEM detector is built by three GEM foils to exploit three multiplication stages. Fig. \ref{gemm} shows a schematic example. The electric fields between the electrodes and the gas mixture also determine the performance of the detector \cite{gem1,gem2}.
The state of the art of the GEMs reports a spatial resolution below 100 $\mu m$ weighting the charge distribution if no magnetic field is present, otherwise the charge information is not enought to measure the position efficiently. A study on another MPGD detector, the MicroMegas, introduced the idea to use the time information to measure the position in presence of low magnetic field \cite{omegas}.
The aim of this work is to demonstrate that the triple GEM technology can achieve spatial resolution of about 100-150 $\mu m$ in presence of high magnetic field. 

\section{Detector geometry and working contributions}
In a triple GEM the five electrodes (cathode, three GEM and anode) define four gaps, namely "conversion" and "drift", between the cathode and the first GEM foil, "transfer 1", "transfer 2", within the GEMs, and "induction" where the signal is induced on the readout plane. The avalanche diffusion and the charge collection depend largely on the value of the electric fields within those gaps. The gain that this detector can achieve is about 10$^4$-10$^5$ \cite{gain}. 
The electrical signal is collected by a segmented anode with two views with 650 $\mu m$ pitch strips. The strips geometry has been optimized to minimize the overlap between the strips \cite{jagged}. The charge is collected by the strips as function of the time and it is used to reconstruct the signal.
Only the electrons generated within the cathode and the first GEM are amplified three times so these electrons dominates the measurement the position of the charge particle.
The gas mixture studied in this work are Argon-CO$_2$ (70:30) and Argon-iC$_4$H$_{10}$ (90:10)\footnote{from now on Argon-CO$_2$ (70:30) and Argon-iC$_4$H$_{10}$ (90:10) will be refered as Argon-CO$_2$ and Argon-iC$_4$H$_{10}$}. These are composed by a streamer (Argon) in order that the multiplication can take place if an high electric field is present, and a quencher (CO$_2$ or iC$_4$H$_{10}$) to reduce the number of seconday electron generated and to prevent the discharge and to operate the detector safely.
An electric field of 3 kV/cm between the GEMs and 5 kV/cm between the anode and the third GEM are used to efficiently extract the signal \cite{gem1} meanwhile in the drift gap is used a value of 1.5 kV/cm. This particular field, named drift field, is responsable of the drift and diffusion properties of the primary electron generated in this region.

 \begin{figure}[!t]
 \centering
 \includegraphics[width=2.5in]{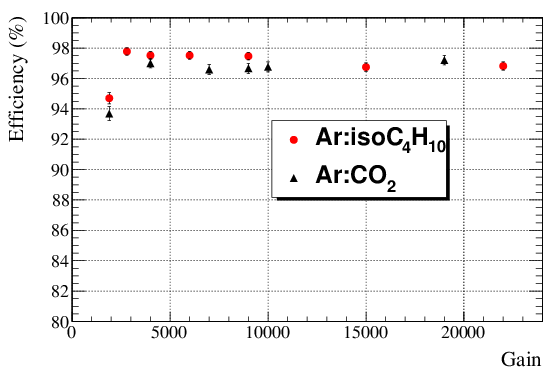}
 \caption{Efficiency measurements of a triple GEM as function of the detector gain. The results show an efficiency plateau above the 95\% for the bidimensional reconstruction for a gain higher that 4000 in both gas mixtures.}
 \label{eff}
 \end{figure}

 \begin{figure}[!t]
 \centering
 \includegraphics[width=2.5in]{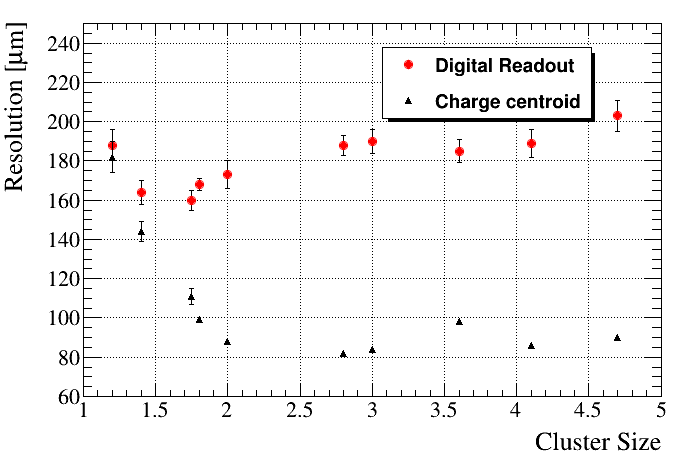}
 \caption{Spatial resolution of an analague (black) and digilat (red) readout as function of the number of fired strip. CC after a certain cluster size reeaches a resolution of 80 $\mu m$ meanwhile the digital readout degrades linearly with the cluster size.}
 \label{digi}
 \end{figure}

 \begin{figure}[!t]
 \centering
 \includegraphics[width=2.5in]{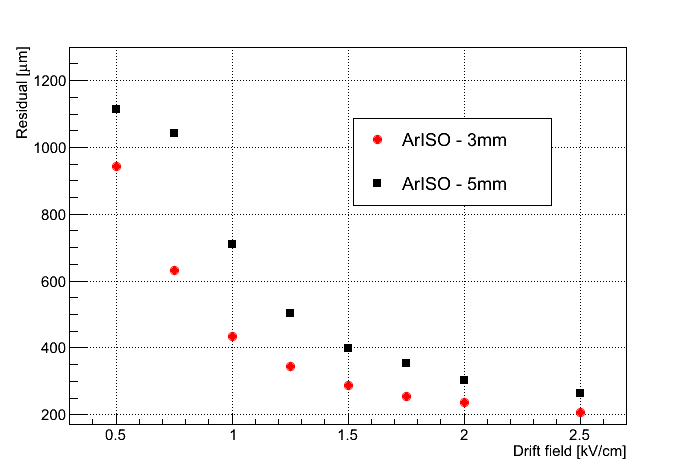}
 \caption{Optimization of the CC for two triple GEM with 3 and 5 mm conversion gap in Argon-iC$_4$H$_{10}$ gas mixture as function of the drift field. A 1 Tesla magnetic field is present. As the field increases the electronic avalanche spread decreasees and the resolution improves up to 190 $\mu m$.}
 \label{ariso}
 \end{figure}

\section{Experimental setup}
Two protoypes of triple GEM with an active area of 10x10 cm$^2$ and with 3 and 5 mm drift gap and 2 mm in the others gaps have been characterized with a muon beam of about 150 GeV/c momentum located at the H4 line at CERN within the RD51 Collaboration \cite{RD51}. These prototypes have been placed within a magnet dipole orthogonal to the beam, Goliath, that can reach 1.5 Tesla in both polarities. A tracking system of four triple GEM, two behind and two below the prototypes, is used to extract the spatial resolutions of the prototypes and the efficiency.
The readout electronic is based on the APV25 hybrid and the Scalar Readout System \cite{APV25} that allow to acquire the information about the charge and the time from each strip by sampling the charge each 25 ns.
Reconstruction algorithms  measure the position throught the clusterization of the contiguous fired strip with a charge above the threshold of 1.5 fC. To measure the position associated to each cluster two algorithm are used.
The Charge Centroid (CC) or center of gravity method exploits the charge values of the readout strips and performe a weighted average of the strip positions and their charge. 
The micro Time Projection Chamber ($\mu$TPC) method exploit the single strip time information to perform a local track reconstruct of the charge particle in few mm drift gap. Each fired strip give two coordinates, one is the strip position and the other (perpendicular to the strip plane) can be reconstructed from the time measurement using the drift velocity of the electron and the Lorentz angle if the magnetic field is present. These values can be extracted from the literature, Garfield simulation or even from direct measurement from the detector \cite{omegas,gas,gem2}. The $\mu$TPC associates to each fired strip a bidimensional point and the ensemble of the strips are lineary fitted to measure the track. The value that correspond to the middle of the gap is associated to the measured position. 


\section{Results}
The voltage difference between the two faces of the GEMs determine the electric field then the detector gain. The first study performed is focused to the research of the operative point, While the gain increase the charge collected increases and therefore the number of fired strips. From experimental data in Fig. \ref{eff} at a gain of 4000 the detector reaches the efficiency plateau for both the studied gas mixtures. Despite to previous reconstruction techniques, as the digital readout, the CC reaches its best performance  if the cluster size is higher than 2.5 (Fig. \ref{digi}). In this configuration, without magnetic field and with orthogonal tracks, the electronic avalanche has a gaussian profile and it is not significantly affected by the diffusion effects. The spatial resolution achieved is 80 $\mu m$. It is not possible to use the $\mu$TPC in this configuration.
\subsection{Reconstruction in magnetic field}
\begin{figure}[!t]
 \centering
 \includegraphics[width=2.5in]{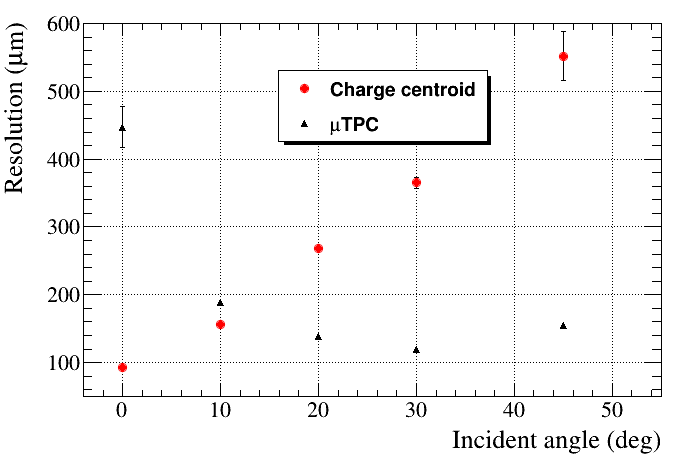}
 \caption{CC and $\mu$TPC spatial resolution as function of the incident angle with Argon-iC$_4$H$_{10}$ gas mixuters. For orthogonal track (0$^\circ$) CC is the best algorithm but as the angle increses the $\mu$TPC reaches 100-150 $\mu m$ spatial resolution. Similar results are obtained with Argon-C0$_2$.}
 \label{b0}
 \end{figure}

The study proceed with the effects of the magnetic field on the performance of the prototypes. As the magnetic field increases, up to 1 Tesla, the Lorentz force acts on the electrons and it increases the dimension of the avalanche. The charge distribution collected at the anode is no more gaussian and this leads to a degradation of the performance of the CC to 300 $\mu m$ in Argon-CO$_2$  and 380 $\mu m$ in Argon-iC$_4$H$_{10}$. 
A strong dependence between the effectiveness of the CC and the drift field has been found since the direction of the electron is described by the Lorentz angle then by the drift field. The performance of the prototypes has been studied as function of this field from 0.5 to 2.5 kV/cm because in this range the Lorentz angle is extremely variable in Argon-iC$_4$H$_{10}$ (Fig. \ref{ariso}). The behavior of the spatial resolution copy the one of the Lorentz angle and gives its best performance at 2.5kV/cm where the angle has a smaller value. At this value it reaches the 190 $\mu m$ with the prototype with 3 mm of drift gap because the diffusion effect are less than the 5 mm prototype.
Each reasult discussed until now has been measured with a beam orthogonal to the prototypes. The introduction of not orthogonal tracks leads to another complication for the CC that degrades as the incidence angle vary from the orthogonal one (Fig. \ref{b0}). The CC gives its best performance if the primary electrons are concentrated around the line that describe their drift to the cathode. As the incident angle increase the region where the primary are generated is larger and the signal collected at the anode is no more gaussian.

 \begin{figure}[!t]
 \centering
 \includegraphics[width=2.5in]{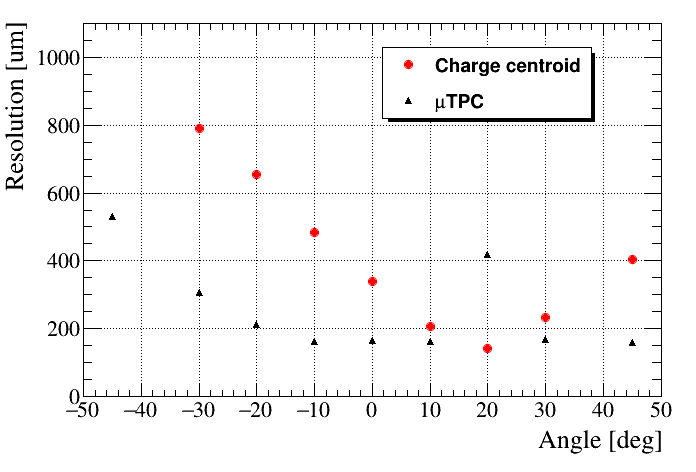}
 \caption{CC and $\mu$TPC as function of the incident angle in 1 Tesla magnetic field and Argon/CO$_2$ gas mixture. Similar results are obtain in Argon-iC$_4$H$_{10}$.}
 \label{b1}
 \end{figure}

 \begin{figure}[!t]
 \centering
 \includegraphics[width=2.5in]{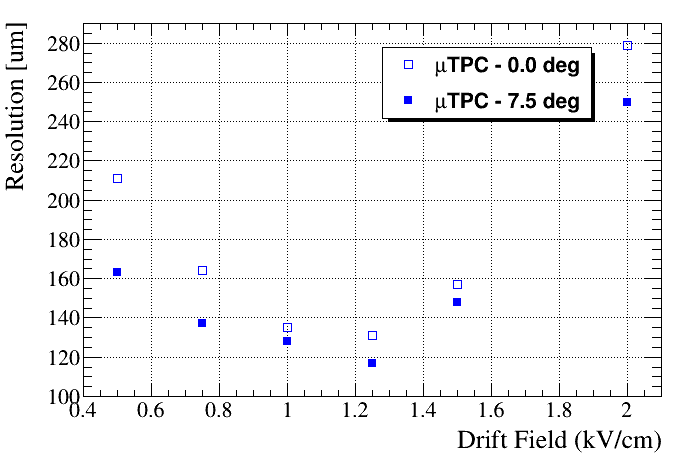}
 \caption{Optimization of the spatial resolution of the $\mu$TPC algorithm as function of the drift field in Argon/CO$_2$.}
 \label{tpc}
 \end{figure}

\subsection{The $\mu$TPC}
This degradation of the CC makes the introduction of the $\mu$TPC of fundamental importance. Although in presence of orthogonal track and no magnetic field this algorithm is not efficent, as the incident angle increases the resolution improves and it reaches a plateu between 100 and 150 $\mu m$. Fig. \ref{b0} shows the experimental results of the CC and $\mu$TPC as function of the angle (0$^\circ$ corresponds to the orthogonal tracks). The two algorithm are anti-correlated and together give a stable performance. The best results have been obtained with the 5 mm drift gap prototype because as the gap is greater as the number of fired strips is higher and this helps the temporal reconstruction. 
Let's examine now the performance results measured in 1 Tesla magnetic field and different incident angle. The Lorentz force gives a preferential direction to the drift of the electron and the combination of the Lorentz angle and the incident angle has to be taken into account.
It is named focusing effect when the Lorentz angle is close to the incident angle because in this configuration the generated primary electrons drift along the same line of the charge particle and the spatial distribution at the anode is concentrate in few strips. If the incident angle departs from the Lorentz one then the defocusing effect take place and counterwise to the focusing effect the electron avalanche is projected on more strips. In the high defocusing region the diffusion effect worsens the $\mu$TPC algorithm. In Fig. \ref{b1} is shown the behavior of the CC and $\mu$TPC as function of the incident angle in magnetic field. The performance are similar at the case without magnetic field but in this case che maximum focusing point coincide with the Lorentz angle: 20$^\circ$ in Argon-CO$_2$, 26$^\circ$ in Argon-iC$_4$H$_{10}$. Here the CC returns its best performance, results comparable to the ones without magnetic field and orthogonal tracks. As the angle departs from this point as the CC degrades and the $\mu$TPC reaches a stable blehavior around 130-150 $\mu m$.
The $\mu$TPC algorithm is the most efficient and as the CC it can be optimize with a drift field study. Conterwise the CC case where the only one parameter to influence it is the Lorentz angle, here it has to be included the dependence from the temporal resolution and the drift velocity \cite{omegas}. A drift field scan has been performed and a value of 1.25kV/cm has been found to improve the $\mu$TPC performance (Fig. \ref{tpc}).

\section{Conclusion}
The performances of a triple GEM in magnetic field have been measured and optimized throught several test beam. The detector amplifies the signal created in the ionization of the gas and measure the signal charge and time with a segmented anode. This information are used to reconstruct the charge particle impact position with two algorithm, CC and $\mu$TPC, that are totally anti-correlated. The combination of these two allow to achieve a stable spatial resolution with result of about 120 $\mu m$ that goes beyond the currect state of the art for this technology in high magnetic field.
The studied technology is of considerable interest in the scientific comunity because its mechanical properties, its electrical stability and the possibility the be shaped the desider form and the studied described in this report can be extended to the large area detector needed in the high energy physics experiments.


%





\ifCLASSOPTIONcaptionsoff
  \newpage
\fi

\end{document}